\documentclass[conference]{IEEEtran}
\IEEEoverridecommandlockouts
\usepackage{cite}
\usepackage{tikz}
\usepackage{amsmath,amssymb,amsfonts}
\usepackage{algorithmic}
\usepackage{graphicx}
\usepackage{textcomp}
\usepackage{xcolor}
\usepackage{soul}
\usepackage{multirow}
\usepackage{threeparttable}
\usepackage{tablefootnote}
\usepackage[numbers]{natbib} %
\usepackage[colorlinks,linkcolor=red,anchorcolor=green,citecolor=blue]{hyperref}
\usepackage{graphicx} %
\def\BibTeX{{\rm B\kern-.05em{\sc i\kern-.025em b}\kern-.08em
    T\kern-.1667em\lower.7ex\hbox{E}\kern-.125emX}}

\newcommand{\chao}[1]{\textcolor{black}{#1}} 
\newcommand{\ryan}[1]{\textcolor{black}{#1}} 
\newcommand{\stef}[1]{\textcolor{black}{#1}}

\newcommand{\mv}[1]{\textcolor{black}{#1}}

\newcommand{\stefnew}[1]{\textcolor{black}{#1}}
\newcommand{\stefnewnew}[1]{\textcolor{black}{#1}}
\newcommand{\stefRoundThree}[1]{\textcolor{black}{#1}}
 
\newcommand{\stefPostSubm}[1]{\textcolor{black}{#1}}

\def\BibTeX{{\rm B\kern-.05em{\sc i\kern-.025em b}\kern-.08em
    T\kern-.1667em\lower.7ex\hbox{E}\kern-.125emX}}
\newcommand\copyrighttext{%
  \footnotesize \textcopyright 2025 IEEE. Personal use of this material is permitted.  Permission from IEEE must be obtained for all other uses, in any current or future media, including reprinting/republishing this material for advertising or promotional purposes, creating new collective works, for resale or redistribution to servers or lists, or reuse of any copyrighted component of this work in other works.
 
  Accepted for publication at the 20th International Conference on PhD Research in Microelectronics and Electronics (PRIME 2025).}

\newcommand{\copyrightnotice}{%
\begin{tikzpicture}[remember picture,overlay]
\node[anchor=south,yshift=10pt] at (current page.south) {\fbox{\parbox{\dimexpr\textwidth-\fboxsep-\fboxrule\relax}{\copyrighttext}}};
\end{tikzpicture}%
}
\begin{document}

\title{iEEG Seizure Detection with a Sparse Hyperdimensional Computing Accelerator\\
}

\author{
    Stef Cuyckens$^\dagger$, Ryan Antonio$^\dagger$, Chao Fang$^{\dagger,\ddagger,*}$, Marian Verhelst$^\dagger$ \\
	\IEEEauthorblockA{
		$^\dagger$KU Leuven, Belgium~~~~~~~~$^\ddagger$Nanjing University, China
    }
    \IEEEauthorblockA{
		Email: \{stef.cuyckens, ryan.antonio, chao.fang, marian.verhelst\}@esat.kuleuven.be
    }
}%
\maketitle
\copyrightnotice %
\vspace{-0.4cm}

\maketitle
\renewcommand{\thefootnote}{}
\footnotetext{$^*$Corresponding author. This project has been partly funded by the European Research Council (ERC) under grant agreement No. 101088865, the European Union’s Horizon 2020 program under grant agreement No. 101070374, the Flanders AI Research Program, and KU Leuven.

Code available at: \url{https://github.com/KULeuven-MICAS/sparse_HDC_for_iEEG}}

\begin{abstract}

\stefPostSubm{Implantable devices for reliable intracranial electroencephalography (iEEG) require efficient, accurate, and real-time detection of seizures. 
Dense hyperdimensional computing (HDC) proves to be efficient over neural networks; however, it still consumes considerable switching power for an ultra-low energy application.}
Sparse HDC, on the other hand, has the potential of further reducing the energy consumption, yet at the expense of having to support more complex operations \stefRoundThree{and introducing an extra hyperparameter, the maximum hypervector density}. 
\stefRoundThree{To improve the energy and area efficiency of the sparse HDC operations, this work introduces the compressed item memory (CompIM) and simplifies the spatial bundling. We also analyze how a proper hyperparameter choice improves the detection delay compared to dense HDC. Ultimately, our optimizations achieve a $1.73\times$ more energy- and $2.20\times$ more area-efficient hardware design than the naive sparse implementation. We are also $7.50\times$ more energy- and $3.24\times$ more area-efficient than the dense HDC implementation. This work highlights the hardware advantages of sparse HDC, demonstrating its potential to enable smaller %
brain implants with a substantially extended battery life compared to the current state-of-the-art.}

\end{abstract}
\vspace{-0.5em}

\section{Introduction} \label{section_intro}

\begin{figure}[t]
    \centering
    \includegraphics[width=1\columnwidth]{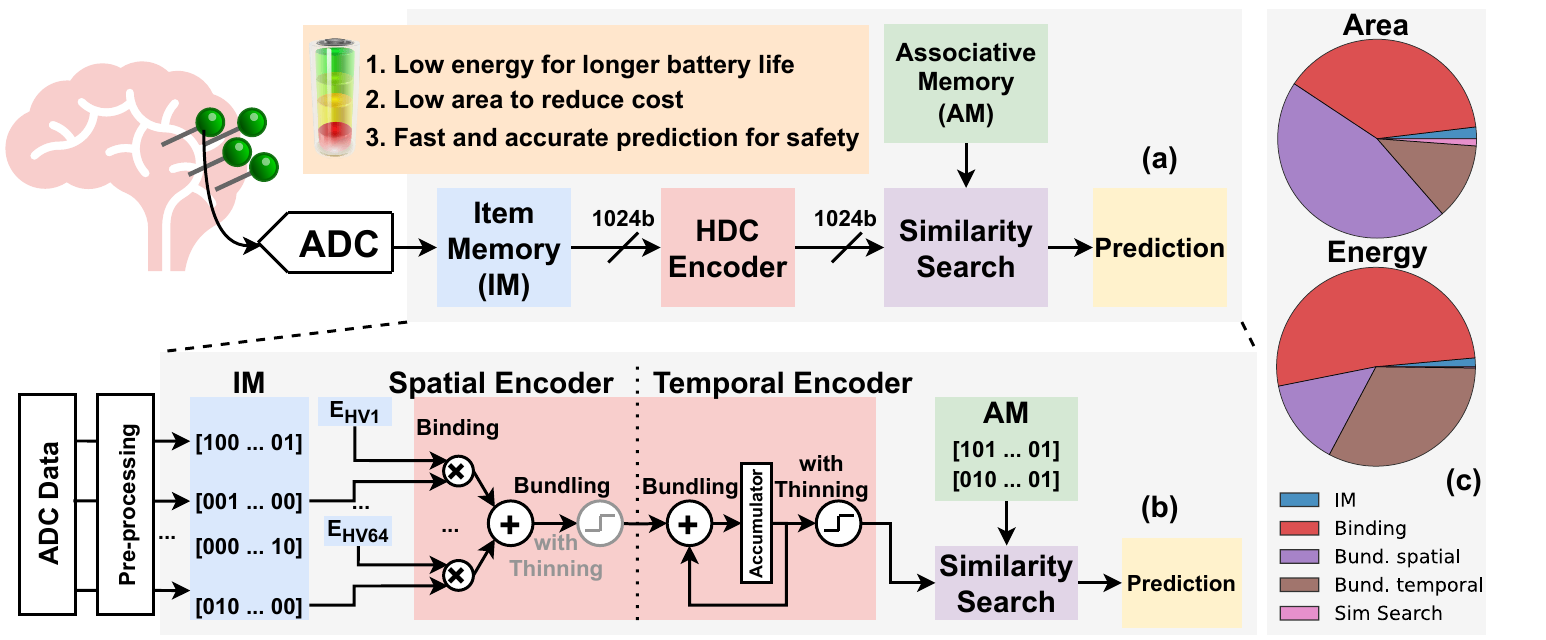}
    \vspace{-2em}
    \caption{(a) High-level overview of iEEG algorithm with HDC and the main requirements, (b) sparse HDC classifier for iEEG system, (c) area and energy breakdown of the naive sparse HDC implementation.}
    \label{fig:introduction}
\end{figure}

The \chao{intracranial electroencephalography} (iEEG) seizure detection application
\ryan{analyzes}
the brain waves of a patient
\ryan{who} 
suffers from epilepsy and predicts if a seizure is about to occur \cite{iEEG_dense_HDC}. %
\ryan{Fig. \ref{fig:introduction}(a) shows that} iEEG seizure \mv{onset} detection requires a computing device and an electrode array \stefPostSubm{attached to} %
the brain. %
iEEG seizure detection \stefPostSubm{requires}:
\stefnew{1) low energy consumption to increase battery lifetime, 2) low production cost realized by lower chip-area requirements,
and 3) fast and reliable prediction of seizures for a safe user experience~\cite{energy_limitation_for_implants, iEEG_dense_HDC}.} %

Recent work has pivoted to \textit{dense} hyperdimensional computing (HDC), thereby avoiding expensive multiply-accumulate (MAC) operations of typical support vector machine (SVM) or neural-network-based implementations \cite{iEEG_dense_HDC}. HDC uses high-dimensional vectors called hypervectors (HV) (with vectorlength $D \geq 1,000$) as representations. These HVs are randomly-generated with a density of 50\% 1-bits and only require efficient element- or bit-wise operations. 
\stefRoundThree{Dense HDC achieves both lower hardware costs and more effective seizure detection compared to neural-network-based state-of-the-art (SotA) implementations \cite{SVM_vs_dense_HDC,iEEG_dense_HDC}}.
\stefRoundThree{Yet, a lot of energy is spent on switching between randomly distributed HVs, due to the high HV-density ($p=50\%$) of dense HDC.} %

\stefPostSubm{Sparse HDC is a promising alternative that uses HVs with 1\% density.}
The sparse nature of the sparse HDC HVs reduces the switching probability of a bit from $50\%$ for dense to only around $2\%$, resulting in very low switching energy.
Sparse HDC uses more complex bit-wise operations than dense HDC. \stefRoundThree{For example,} dense binding is implemented by a simple bit-wise XOR, while for sparse HDC, the segmented shift binding requires circularly shifting the segments of one input HV by the positions of the 1's in the corresponding segments of the other input HV \cite{Schlegel_Neubert_Protzel_2021}. %
\stefRoundThree{These more complex bit-wise operations necessitate thorough optimization of the classification system, as computational overhead could otherwise counteract the energy efficiency benefits inherently provided by the sparse HDC representation.}

\stefRoundThree{To \stefRoundThree{optimize} %
sparse HDC, a baseline implementation of the sparse HDC classifier in Fig. \ref{fig:introduction}(b) is analyzed and broken down by module in Fig. \ref{fig:introduction}(c). By relying on this breakdown, we target binding and spatial bundling with our two proposed optimizations.}
(1) We develop the compressed item memory to simplify the binding by leveraging the sparsity in the item memory. (2) We simplify the spatial bundling by removing the thinning. These reduce the dominant energy and area costs.

\stefRoundThree{In summary, t}his work seeks to optimize the hardware (HW) implementation of sparse HDC for the iEEG seizure detection application. To the best of our knowledge, we are the first to develop a HW implementation of sparse HDC for an application with many input channels.
\mv{While t}he optimizations \mv{are applied to iEEG applications in} this paper, \mv{they} can be reused for other sparse HDC applications \mv{with many input channels, such as} voice recognition \cite{voice_recog}, DNA sequencing \cite{dna_seq}, and other biomedical sensor applications \cite{other_bio_apps}.

\section{\mv{Sparse HDC f}undamentals} \label{section_fundamentals}

\stefRoundThree{The sparse HDC classification system was adapted from the dense HDC classification system for iEEG seizure detection from \cite{iEEG_dense_HDC} by changing the dense HDC operations to their sparse equivalents. The system is split into 5 parts/modules, as shown in Fig.~\ref{fig:introduction}(b): the IM, the many parallel bindings, the bundling with thinning in the spatial encoder, the bundling with thinning in the temporal encoder, and the similarity search with the associative memory (AM). 
Before going to the HDC classifier, the electrode data is preprocessed into 6-bit local binary pattern codes (LBP codes), which capture the relation between consecutive values and are used by the SotA iEEG system with dense HDC in \cite{iEEG_dense_HDC}.}

\subsection{Item Memory (IM)}

The IM maps the input data to an HV in HDC space for each electrode. 
Each of the 64 electrodes produces an LBP code each clock cycle, the IM maps each of these LBP codes to a 1024-bit sparse HV with 8 1-bits ($p \approx 1\%$) using a look-up table (LUT). This requires a LUT for each channel/electrode, but because the HVs in the IM are randomly generated at design time and only a few bits are 1's for each HV, the IM can be heavily optimized by the design tools.

\subsection{Binding}

The binding ($\bigotimes$) operation combines two HVs: one representing the sensor data from IM and another representing the electrode that collected this data. This combination creates a unique output HV that preserves both what the data is and which electrode detected it, enabling the system to maintain electrode-specific information throughout processing \cite{Schlegel_Neubert_Protzel_2021}.

\begin{figure}[t]
    \centering
    \includegraphics[width=1\columnwidth]{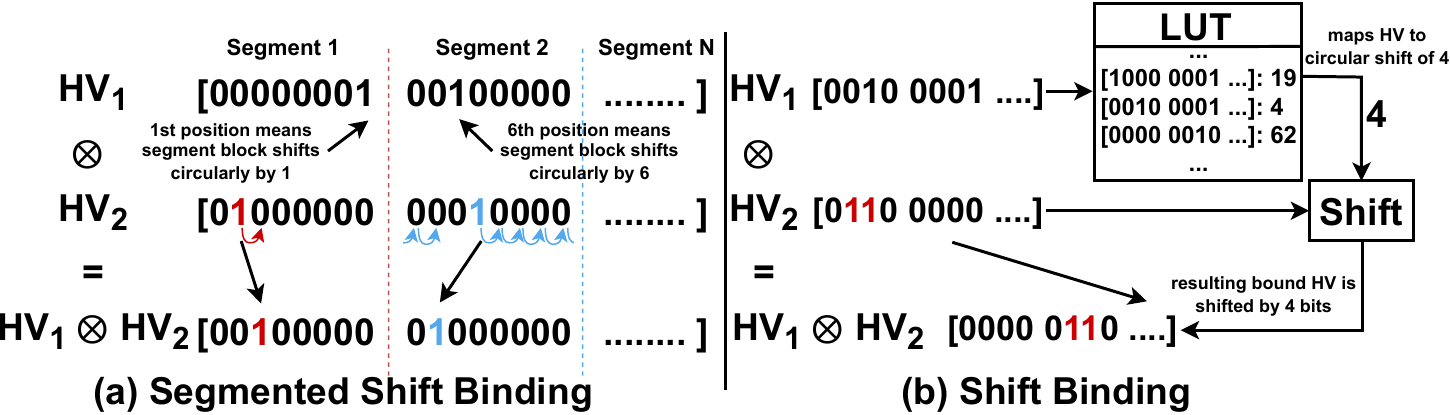}
    \vspace{-2em}
    \caption{(a) Segmented shift binding and (b) shift binding.}
    \label{fig:segm_shift_bind}
\end{figure}

\stefPostSubm{There are} two binding operations \cite{lang_recog}, the segmented shift binding and the shift binding \cite{laiho,Schlegel_Neubert_Protzel_2021}. Fig. \ref{fig:segm_shift_bind}(a) illustrates the segmented shift binding. The segmented shift binding splits the 1024-bit HV in 8 equally long segments, each containing a single 1-bit and 127 0-bits. The binding involves circularly shifting each of the first input HV segments by the position of the 1-bit in the corresponding segment of the other input HV. Fig. \ref{fig:segm_shift_bind}(b) illustrates the shift binding. The shift binding maps one input HV to an integer by a LUT, the other input HV is then shifted by that integer. This requires a large LUT that maps 1024-bit input HVs to integers \cite{Schlegel_Neubert_Protzel_2021}. Due to the large area cost associated with this large LUT for the shift binding, we do not consider the shift binding further and focus our implementation on the segmented shift binding.

The \stefRoundThree{naive} implementation of the segmented shift binding is illustrated in Fig. \ref{fig:naiv_vs_opt_encoder}(a), where we first extract the positions of the 1-bits from the data-representing HVs that come out of the IM. The 1-bit positions are computed by a one-hot to binary decoder for each segment of the HV. Then, the electrode-representing HV segments are shifted by those positions in a barrel shifter.

\subsection{Bundling with Thinning} \label{section-2c}

The bundling ($+$) combines many HVs to collect their information in one HV. This increases the density of the HV, which could lead to HVs that are completely filled with 1-bits when many sparse HVs are combined. No proper classification can be performed with completely filled HVs, so to circumvent this, the bundling is often paired with a thinning \cite{Schlegel_Neubert_Protzel_2021}. This is also the case in the iEEG HDC algorithm \cite{iEEG_dense_HDC} for both the bundling in the spatial encoder and the bundling in the temporal encoder, as seen in Fig. \ref{fig:introduction}(b). \stefRoundThree{The baseline implementation of the spatial bundling with thinning consists of}
an adder tree for each element of the HV, which adds up all the corresponding bits in the 64 bound HVs. Then, a threshold is applied to thin the result back to a binary representation. Fig. \ref{fig:naiv_vs_opt_encoder}(a) illustrates the implementation of the spatial bundling. The threshold decides the density of the output HV and should be chosen to balance the algorithmic performance with the switching energy. The density after bundling is typically much larger than the density of the original sparse HVs \cite{lang_recog}.

The temporal bundling is implemented similarly. In the temporal encoder, the bundling combines 256 sequential output HVs of the spatial bundling \cite{iEEG_dense_HDC}. The HVs are added up and thinned with a threshold, and the intermediate result is saved in a large register that is \stefRoundThree{8 bits} long for each of the HV elements, in total a large 8192-bit register. The temporal bundling gives an output every 256 clock cycles, we call this a time frame.

\subsection{Similarity Search with AM}
In the similarity search, the output HVs of the temporal bundling are compared with the class-representing HVs stored in the AM. These class-representing HVs were computed through the same sparse HDC classifier as the inference but with labeled data from one seizure. To combine the data from multiple time frames, we do an additional bundling when training with thinning to 50\% density. This combines all the time frames with the same class in a single class-representing HV \cite{iEEG_dense_HDC}. %
The training is implemented offline.

The comparison is implemented by an AND-gate for each element of the HVs and an adder tree that adds the output elements of the AND-gates. This forms a similarity metric that only looks at the 1-bits of the HVs, as there is no information in the 0-bits for sparse HDC \cite{Schlegel_Neubert_Protzel_2021}. This is done sequentially for the two classes, then the similarity scores are compared, and the highest one decides the classification.

\section{\stefRoundThree{Sparse HDC Optimizations}} \label{section_imp}

\stefRoundThree{To optimize the energy and area of the sparse HDC classification system, we target the dominant modules in the energy and area breakdowns of Fig. \ref{fig:introduction}(c). We target the binding to primarily increase the energy-efficiency by developing the CompIM. To reduce the area, we simplify the spatial bundling by removing the thinning.}

\subsection{Compressed IM (CompIM)} \label{section_impA}

\begin{figure}[t]
    \centering
    \includegraphics[width=0.9\columnwidth]{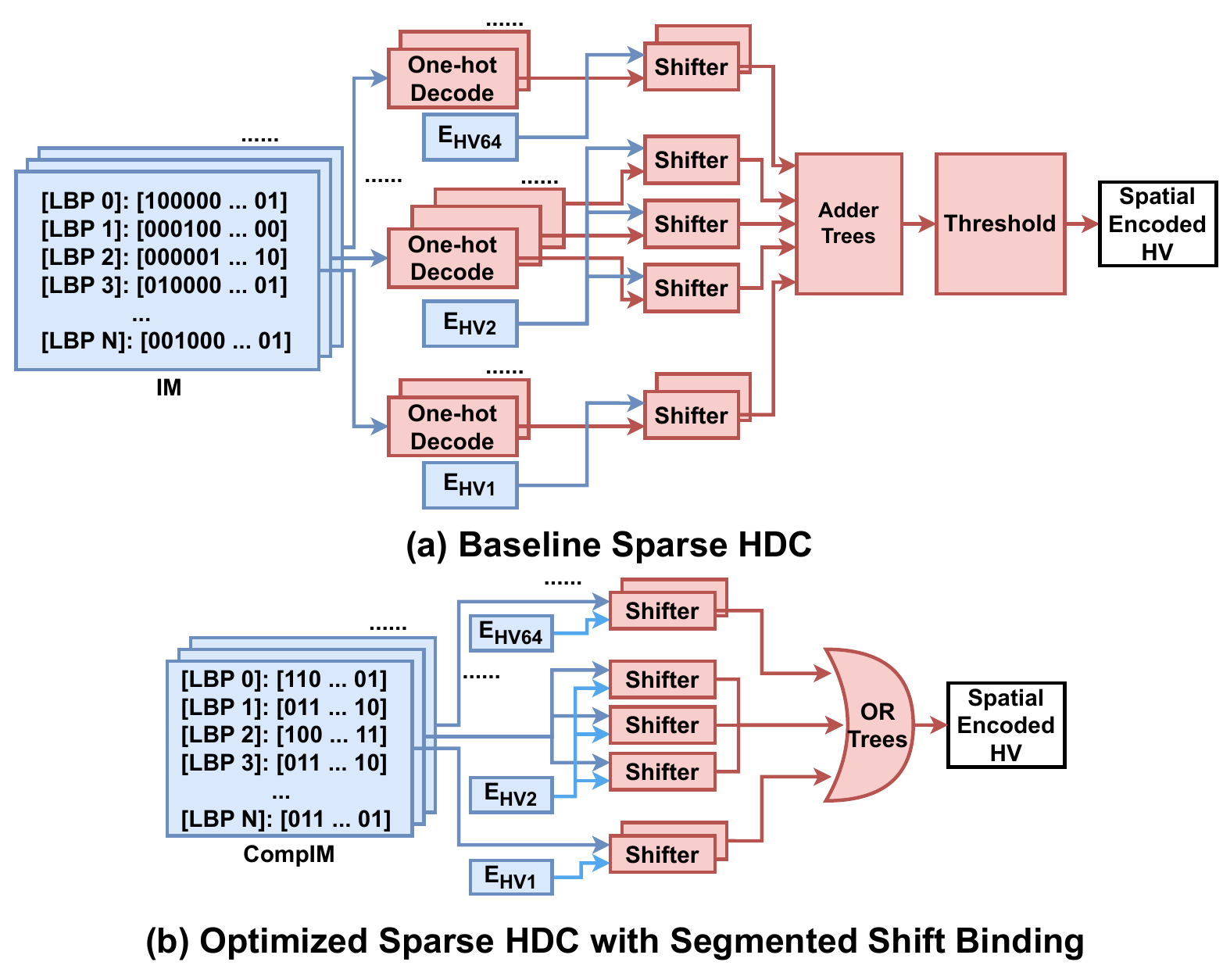}
    \vspace{-1em}
    \caption{Overview of HW implementation of first part of baseline sparse HDC system (a) and of optimized sparse HDC system (b).}
    \label{fig:naiv_vs_opt_encoder}
\end{figure}

Fig. \ref{fig:introduction}(c) shows the energy and area breakdown of the baseline sparse HDC classifier. Here, we see that the binding together with its one-hot decoder takes up 51.3\% of the energy and 38\% of the area. We seek to reduce the area and energy of the binding by integrating the one-hot decoder with the IM, we call this the compressed IM (CompIM). 
Now, instead of transforming the input data to 1024-bit HVs, the CompIM links the input data directly to the positions of the 1-bit in each segment. This allows the one-hot decoding to be removed from the binding as this is now compressed into the CompIM. This reduces the number of bits to represent a HV in the IM from 1024 to \stefPostSubm{just 56 bits (8 segments × 7 bits for position in the segment)}.
The architecture changes are illustrated in Fig. \ref{fig:naiv_vs_opt_encoder}.

\stefnewnew{We can combine the one-hot decoding step in the IM, as unlike dense HDC where the information is distributed over the whole HV, sparse HDC only keeps information in the positions of the few 1-bits. By leveraging this sparsity, we compress the one-hot decoding into the sparse IM. This compression makes the CompIM dense, albeit much smaller than the IM of dense HDC. The CompIM uses the sparsity in the IM to simplify the binding operation, the energy and area benefits of this optimization are explored in Sec.~\ref{section_experimentsB}.}

\subsection{Spatial Bundling Without Thinning} \label{section_impB}

The area breakdown of the baseline sparse HDC system in Fig. \ref{fig:introduction} shows that an overwhelming 44.9\% of area goes to the spatial bundling. This subsection seeks to reduce this to get a more cost-effective design.

The baseline implementation, explained in Sec.~\ref{section_fundamentals}, was adapted from the dense HDC classifier for iEEG seizure detection of \cite{iEEG_dense_HDC} by transforming the dense HDC operations to their sparse equivalents. This same technique was used in \cite{Schlegel_Neubert_Protzel_2021} to compare different HDC variants. Now, the dense HDC bundling is always followed by a thinning \cite{iEEG_dense_HDC,Schlegel_Neubert_Protzel_2021,SVM_vs_dense_HDC}, but this is not always necessary for sparse HDC \cite{Schlegel_Neubert_Protzel_2021}. 

As mentioned in Sec.~\ref{section_fundamentals}, thinning makes sure that the HVs do not become completely filled. In our implementation of the iEEG sparse HDC classifier, the spatial bundling combines 64 HVs with a density of 0.78\%. This comes down to a maximum density after the spatial bundling of 50\% if no overlap occurs. The HV cannot get filled entirely in this first bundling step, so thinning can be removed. However, this could affect the algorithmic performance as this does change the output of the spatial bundling. The algorithmic performance with and without thinning on the spatial bundling is explored in Sec.~\ref{section_experimentsA}. 

Removing the thinning allows the spatial bundling to be simplified from adder trees to OR trees, as shown in Fig. \ref{fig:naiv_vs_opt_encoder}(b). The effect on the energy and area is explored in Sec.~\ref{section_experimentsB}.

\section{Experiments} \label{section_experiments}

\mv{To assess the benefits of the proposed innovations, we will first}
\mv{assess the accuracy impact of our proposed optimizations, followed by an elaborate analysis of their impact on hardware efficiency.}
\stef{Specifically, we compare the \mv{proposed designs of} %
Sec.~\ref{section_imp} with \mv{a baseline} sparse and dense implementation of the iEEG system, in which we break down the energy and area by module.
Finally, we compare our sparse HDC system with two SotA iEEG seizure detection implementations and one similar dense HDC application.
All experimental results are obtained from synthesis in a TSMC 16nm FinFET technology. The energy analysis is obtained using Synopsys PrimeTime PX with switching annotations. \stefPostSubm{We use the one-shot learning subset of the iEEG seizure detection dataset from \cite{iEEG_dense_HDC}. The energy and area analysis were carried out on seizure data from patient 11 of \cite{iEEG_dense_HDC}.}}

\begin{table}[b]
    \caption{Comparison to SotA}
    \label{tbl:SotA}
    \centering
    \resizebox{\columnwidth}{!}{ 
        \begin{tabular}{|c|c|c|c|c|}
            \hline
            Specs                         & \textbf{Ours*} & \cite{SVM}* & \cite{OLeary} & \cite{SVM_vs_dense_HDC}       \\ \hline \hline
            Application                   & \begin{tabular}[c]{@{}c@{}} iEEG Seizure \\  Detection \end{tabular}    & \begin{tabular}[c]{@{}c@{}} EEG Seizure \\  Detection \end{tabular}  &  \begin{tabular}[c]{@{}c@{}} iEEG Brain \\  State Classification \end{tabular} & \begin{tabular}[c]{@{}c@{}} Emotion \\ Recognition \end{tabular}   \\ \hline
            Type                          & Sparse HDC    & SVM    &  Decision Tree  & Dense HDC   \\ \hline
            Technology (nm)               & 16            & 65      &  65    & 28          \\ \hline
            Voltage (V)                   & 0.75          & -     &  1.2   & 0.8         \\ \hline
            Frequency (MHz)               & 10            & 100       &  -     & 0.909       \\ \hline
            HV Dimension                  & 1,024         & -       &  -     & 2,000       \\ \hline
            Channels                      & 64            & 23      & 8      & 214         \\ \hline
            Area ($mm^2$)                 & 0.059         & 0.09       &  1.95  & 0.068       \\ \hline
            Latency per predict           & 25.6 $\mu s$  & 160 $ns$   &  -     & 1 $ms$      \\ \hline
            Energy per predict ($nJ$)      & 12.5          & 841.6    &  36    & 39.1        \\ \hline
            Energy/Channel ($nJ$)         & 0.195         & 36.59    &  4.5   & 0.183        \\ \hline
        \end{tabular}
    }

    \begin{tablenotes}
        \begin{scriptsize}
            \item[*] * Synthesized results only.
        \end{scriptsize}
    \end{tablenotes}
\end{table}

\subsection{%
\stefPostSubm{Assessment of Algorithmic Performance}
} \label{section_experimentsA}

We evaluate our sparse HDC implementation on algorithmic level by one-shot learning on a seizure and testing on all other seizures of a patient. 
\stefPostSubm{We compare with the dense HDC baseline of \cite{iEEG_dense_HDC} on two metrics: the detection delay and the \stefPostSubm{seizure detection accuracy}, both averaged over the one-shot learning patients of \cite{iEEG_dense_HDC}.}
The detection delay is found by the delay from the seizure onset point marked by an expert \cite{iEEG_dense_HDC}. This seizure onset often precedes physical/clinical symptoms by more than 20 seconds \cite{iEEG_dense_HDC}. Our objective is to \stefPostSubm{maximize detection accuracy and} minimize detection delay, as a shorter time interval between alert and seizure symptoms leads to a safer and more favorable user experience. 

\begin{figure}[t]
    \centering
    \includegraphics[width=0.95\columnwidth]{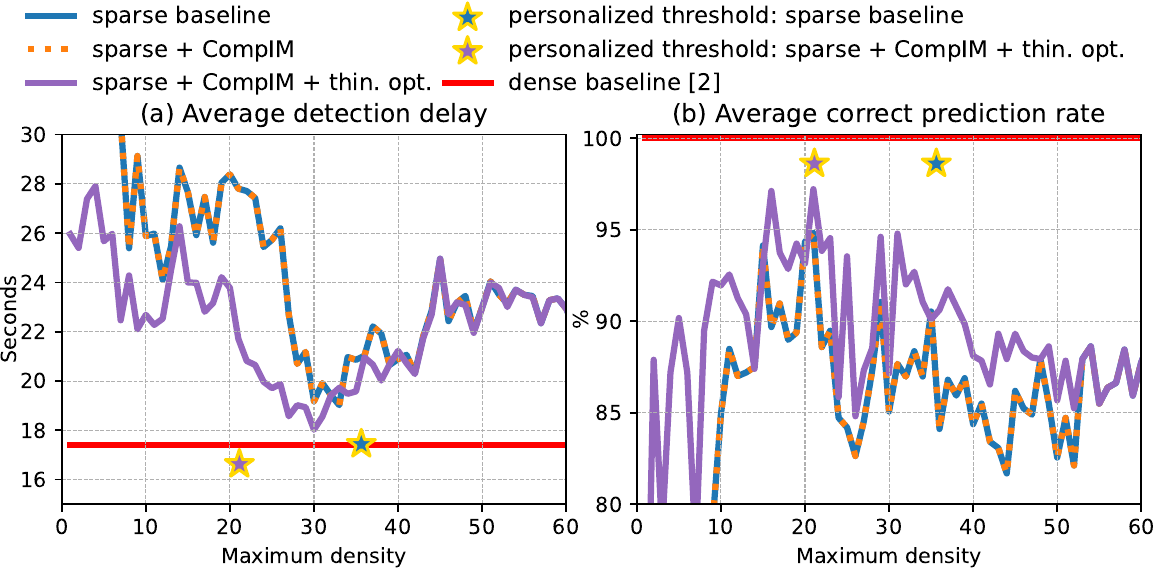}
    \vspace{-1em}
    \caption{Average seizure detection delay and detection accuracy for different maximum HV densities after bundling.}
    \label{max_density_vs_detec_delay}
    \vspace{-1em}
\end{figure}

Fig. \ref{max_density_vs_detec_delay} compares the detection delay and detection accuracy of the sparse and dense baseline implementations with our optimizations. Sparse HDC has an extra hyperparameter, the maximum HV density after thinning, which should be tuned 
to get a low detection delay and a high detection accuracy. \stefPostSubm{The lines show the average values when each patient has the same maximum density, but this assumes that the maximum density is tuned over all patients. \stefPostSubm{In practice, the hyperparameter is tuned} for each patient individually; the stars in Fig. \ref{max_density_vs_detec_delay} indicate the best performance possible when each patient has their optimal maximum density.
If the maximum density hyperparameter is tuned, sparse HDC can achieve a lower detection delay, but falls short of the 100\% detection accuracy of dense HDC.}

\vspace{-0.5em}
\subsection{Energy and Area Breakdown} \label{section_experimentsB}

\begin{figure}[t]
    \centering
    \includegraphics[width=0.95\columnwidth]{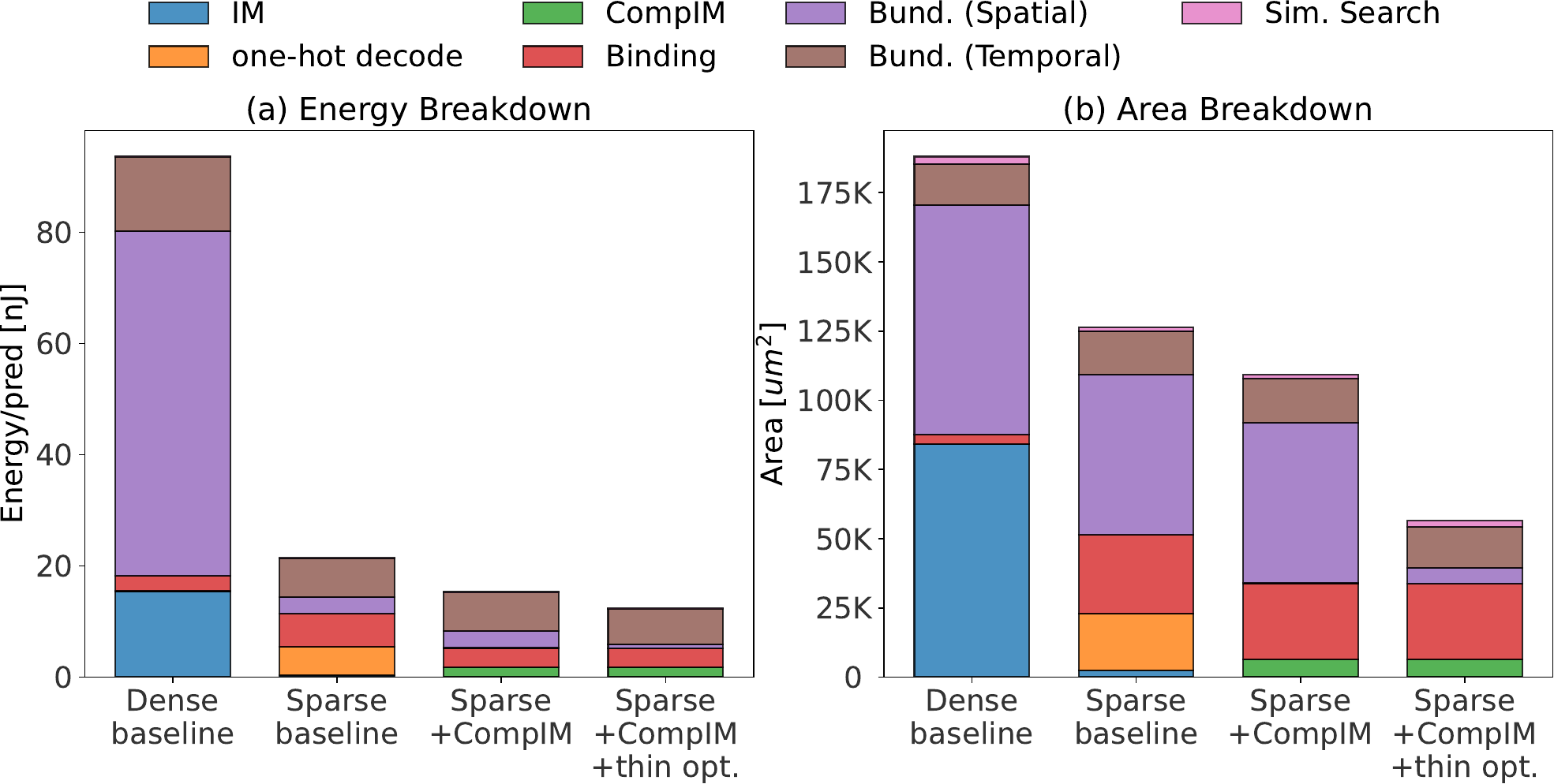}
    \vspace{-1em}
    \caption{Energy and area breakdown of the iEEG system, comparing dense and sparse HDC baselines with our optimizations.}
    \label{area_power_breakdown}
    \vspace{-0.3em}
\end{figure}

We evaluate our implementation and the baselines with a clock frequency of 10MHz \stefRoundThree{and a threshold of 130 to keep the maximum HV density after thinning between $20-30\%$}\stefPostSubm{, in line with the purple star in Fig. \ref{max_density_vs_detec_delay}}. Fig. \ref{area_power_breakdown} compares our optimizations with the sparse and dense HDC baseline implementations. Our CompIM combines the IM and one-hot decoding and reduces the overall energy and area overhead. Our thinning optimization reduces the area considerably. \stefRoundThree{In total, we achieve a $1.72\times$ higher energy efficiency and a $2.20\times$ higher area efficiency compared to the sparse HDC baseline. Compared to the dense HDC baseline, our implementation is $7.50\times$ more energy- and $3.24\times$ more area-efficient.}

\vspace{-0.5em}
\subsection{SotA Comparison}

We compare our optimized sparse HDC implementation with an SVM \cite{SVM} and decision tree \cite{OLeary} implementation of the seizure detection application. We also compare with a similar dense HDC application. Table \ref{tbl:SotA} shows that our work is the most energy- and area-efficient \stefRoundThree{among these SotA prior arts}. \stefnewnew{We use bit-wise operations and high sparsity, which allows us to be more efficient than \cite{SVM} and \cite{OLeary}. 
On the other hand, when normalizing over the number of channels, our energy efficiency becomes comparable to \cite{SVM_vs_dense_HDC}, despite them using dense HDC. This is the case because their emotion recognition uses a different HDC algorithm, where the temporal encoder only operates once per prediction, while ours runs 256 times per prediction. As a result, we process $64 \times 256$ HVs per prediction, compared to just 214 HVs in their case, accounting for the close energy/prediction values.}

\section{Conclusion}

This work introduces two hardware optimizations for iEEG seizure detection with sparse HDC using the segmented shift binding. We combine the IM with the one-hot decoding of the binding into the CompIM. We analyse the effect of thinning on the spatial bundling and conclude that the adder trees in this step can be reduced to OR trees without algorithmic performance loss. This results in a $1.72\times$ energy and $2.20\times$ area reduction compared to the sparse HDC baseline. 
\stefPostSubm{Furthermore, with proper selection of the maximum density after thinning hyperparameter, the sparse HDC classifier can achieve similar algorithmic performance to dense HDC. Ultimately, our research pushes the efficiency of sparse HDC further and shows the viability of sparse HDC classification 
to enable smaller brain implants with improved battery life.}

\bibliographystyle{IEEEtran}

{\footnotesize
\bibliography{refs}
}

\end{document}